\documentclass{emulateapj}
\usepackage{natbib}

\slugcomment{To appear in the AJ}
\shorttitle{SN 2008bk Echo}
\shortauthors{Van Dyk}

\begin{document}

\title{An Echo of Supernova 2008bk\footnote{Based in
part on observations with the NASA/ESA {\sl Hubble Space Telescope}, obtained
at the Space Telescope Science Institute (STScI), which is operated by AURA,
Inc., under NASA contract NAS5-26555.}}

\author{Schuyler D.~Van Dyk\altaffilmark{1}}

\altaffiltext{1}{Spitzer Science Center/Caltech, Mailcode 220-6,
  Pasadena CA 91125; email: vandyk@ipac.caltech.edu.}

\begin{abstract}
I have discovered a prominent light echo around the low-luminosity Type II-Plateau Supernova (SN)
2008bk in NGC 7793, seen in archival images obtained with the Wide Field 
Channel of the Advanced Camera for Surveys on-board the {\sl Hubble
Space Telescope\/} ({\sl HST}).  The echo is a partial ring, brighter to the north and east than to 
the south and west. The analysis of the echo I present
suggests that it is due to the SN light pulse scattered by a sheet, or sheets, of dust
located $\approx$15 pc from the SN. The composition of the dust is assumed to be of 
standard Galactic diffuse interstellar grains.
The visual extinction of the dust responsible for the echo is $A_V \approx 0.05$ mag, 
in addition to the extinction due to the Galactic foreground toward the host galaxy.
That the SN experienced much less overall extinction implies that it is seen through a less
dense portion of the interstellar medium in its environment.
The late-time {\sl HST\/} photometry of SN 2008bk also clearly demonstrates that the progenitor star 
has vanished.
\end{abstract}


\section{Introduction}

A light echo from a supernova (SN) explosion is likely to be a common occurrence.  
When not resolved, the presence of light echoes around extragalactic
supernovae (SNe) 
can be inferred based on excesses in their late-time optical and infrared light curves
\citep[e.g.,][]{wright80,dwek83,graham83,graham86,schaefer87,milne03,welch07,mattila08a,miller10,meikle11,otsuka12,sugerman12}.
Recently, the detections of ancient light echoes around SNe in the 
Galaxy \citep{krause05,krause08a,krause08b,rest08b}
and the Large Magellanic Cloud \citep[LMC;][]{rest05,rest08a} have spectacularly realized 
what both \citet{shklovskii64} and \citet{vdbergh65,vdbergh66} had  
speculated about decades ago.
Echoes can also reveal the properties of energetic Galactic events that are not quite (yet)
SNe \citep[e.g.,][]{bond03,rest12a}.

Up till now, 
eight nearby extragalactic SNe have had spatially-resolved echoes discovered around 
them: SN 1987A in the LMC \citep[e.g.,][]{crotts88,emmering89,bond90}, 
SN 1991T in NGC 4527 \citep{schmidt94,sparks99}, 
SN 1993J in Messier 81 \citep{sugerman02,liu03}, 
SN 1995E in NGC 2441\citep{quinn06}, 
SN 1998bu in Messier 96 \citep{cappellaro01},
SN 1999ev in NGC 4274 \citep{maund05},
SN 2003gd in Messier 74 \citep{sugerman05,vandyk06}, 
and
SN 2006X in Messier 100 \citep{wang08,crotts08}.
Although both the interstellar and circumstellar echoes from the Type II SN 1987A could
be discovered from the ground, the echoes around the other SNe 
had to be revealed by the superior angular 
resolution of the {\sl Hubble Space Telescope\/} ({\it HST}).

These light echoes result from the luminous ultraviolet (UV)/optical emission
pulse from a SN being scattered by dust in 
the SN environment.
The UV pulse will tend to photoionize much of the circumstellar matter and destroy
smaller dust grains nearest to the SN, while more distant, larger grains
survive the pulse. 
The light echo we see at a given instant is the intersection of the dust filament or sheet with the 
(virtual) ellipsoid surface of constant arrival time associated with that particular instant of time. 
Such an intersection is in nearly all cases a circle or arclet.
Light echoes can probe both the circumstellar and the largest interstellar structures
in the SN environment \citep{sugerman03}.
From the scattered SN light, the size distribution and
composition of the dust grains can be determined \citep[e.g.,][]{sugerman03}.
Light echoes also provide a means to measure the distance to the
SN, and therefore its host galaxy, 
based purely on geometrical arguments and independent of any distance
ladder \citep{panagia91,sparks94,sparks96}.
Light echoes also allow three-dimensional spectroscopy of transients \citep[e.g.,][]{rest12b}.

\bibpunct[; ]{(}{)}{;}{a}{}{;}
I have discovered a resolved light echo around the SN II-Plateau (II-P) 
2008bk in the nearby host galaxy NGC 7793. 
SN 2008bk has been recently analyzed by 
\citet[][also \citeauthor{mat08b}~\citeyear{mat08b}]{van12a}. 
Furthermore, G.~Pignata et al.~(in preparation) possess a rich dataset on this SN. 
In these studies the spectroscopic and photometric properties of
the SN have been shown to most closely resemble the low-luminosity SN II-P 1999br in NGC 4900 
\citep{pas04}. Furthermore, \citeauthor{van12a} characterized the nature of the
red supergiant (RSG) progenitor star of SN 2008bk, seen in very high-quality, multi-band
ground-based images obtained prior to explosion. 
Those authors were able to constrain
the initial mass of the progenitor to 8--8.5 $M_{\odot}$.
The presence of the echo around SN 2008bk has been revealed through 
very late-time observations of the host galaxy with {\sl HST}.

\section{Observations}\label{observations}

\subsection{{\sl HST\/} Data}

The site of SN 2008bk was observed on 2011 January 17.23 UT
with the Advanced Camera for Surveys
(ACS) Wide Field Channel (WFC) as part of program GO-12196 (PI: D.~Radburn-Smith).  
The explosion epoch was likely $\sim$ JD 2454548 (G.~Pignata et al., in preparation), 
so these images were obtained when the SN was at an age $\sim 1024.7$ d, or $\sim$2.81 yr.
The bandpasses
and exposure times were F606W (999 s) and F814W (1175 s). 
The site was also observed on 2011 April 30.0, at SN age $\sim$3.11 yr, 
with both the UVIS and IR channels of
the {\sl HST\/} Wide-Field Camera 3 in F814W (915 s), F125W (461.8 s), and F160W (828.6 s),
all in subarray mode, as part of GO-12262 (PI: J.~Maund). 
I obtained all of these
data from the {\sl HST\/} public archive, where standard STScI pipeline procedures
had been employed to calibrate the images.
The ACS/WFC data had been corrected for charge-transfer efficiency (CTE) losses in the pipeline.

The F606W ($\sim V$) image from 2011 January 
is shown in Figure~\ref{figecho}.
The field is crowded and replete in stars of various brightnesses
around the SN position within the host galaxy.
The prominent light echo can clearly be seen in the {\sl HST\/} image.
The object inside the echo is SN 2008bk. The light from the SN is presumably responsible for the
echo. The echo is also detected in the ACS and
WFC3 F814W images, but not in the WFC3/IR images.

As seen most clearly at F606W, the light echo is an incomplete, although notably quite circular, arc of 
emission, most prominent toward the north and far less evident in the south.
The echo is less prominent in the F814W images. 
The echo center does not exactly coincide with the SN position, which indicates that the dust
distribution is itself inclined with respect to our line-of-sight.
(\citeauthor{tylenda04}~\citeyear{tylenda04} showed that, even for an inclined dust sheet, the
shape of a light echo is still approximately circular, although with the center of the echo offset from the
source.)
The point-spread function (PSF) of the SN appears more extended due east in the F555W image, 
possibly due to the presence of a fainter, blue star; the profile is less extended in the F814W image.

I measured the SN brightness in both F606W and F814W, using the package
Dolphot v2.0 \citep{dolphin00}, with the ACS package, 
which automatically
accounts for WFC PSF variations across the chips, 
zeropoints, aperture corrections, etc.
The output from the package automatically includes the transformation from flight-system F606W
and F814W to the corresponding Johnson-Cousins \citep{bessell90} magnitudes (in $V$ and $I$), 
following the prescriptions of \citet{sirianni05}. 
I also analyzed the WFC3 subarray data using Dolphot.
The resulting photometry for the SN from all of these images is given in Table~\ref{tabphot}.

The brightness of the SN in 2011 January was 
m$_{\rm F606W}$=23.284 ($V$=23.226) $\pm 0.011$ and
m$_{\rm F814W}$=23.767 ($I$=23.762) $\pm 0.025$ mag, which are 
0.5 mag and 3.0 mag fainter in $V$ and $I$, respectively, 
than the observed brightness of the progenitor RSG \citep{van12a}.
No doubt should therefore exist that this star was seen to explode in 2008 and has subsequently
vanished. (See also \citeauthor{mattila10}~\citeyear{mattila10}.)

\subsection{Ground-Based Data}

I obtained publicly available image data from the European Southern Observatory (ESO) archive. 
The images were acquired in
$V$ and $I$ bands at the 3.58 m New Technology Telescope (NTT) with
the ESO Faint Object Spectrograph and Camera (EFOSC2) 
as part of programs 082.A-0526 (PI: M.~Hamuy), 
083.D-0970 and 184.D-1140 (PI of both: S.~Benetti), as well as in $V$ only at the 
8.2 m Very Large Telescope (VLT) 
Unit Telescope~1 with the FOcal Reducer and low dispersion Spectrograph (FORS2) 
as part of program 083.D-0131 (PI: S.~Smartt).
I used PSF fitting of these
images using DAOPHOT \citep{stetson87} within IRAF\footnote{IRAF (Image Reduction and 
Analysis Facility) is distributed by the National Optical Astronomy Observatories, which are 
operated by the Association of Universities for Research in Astronomy, Inc., under
cooperative agreement with the National Science Foundation.} to extract photometry of the SN and 
calibrated the 
photometry using the local stellar sequence shown in \cite{van12a}. The results of this photometry
are given in Table~\ref{tabphot}.

\begin{figure}
\plotone{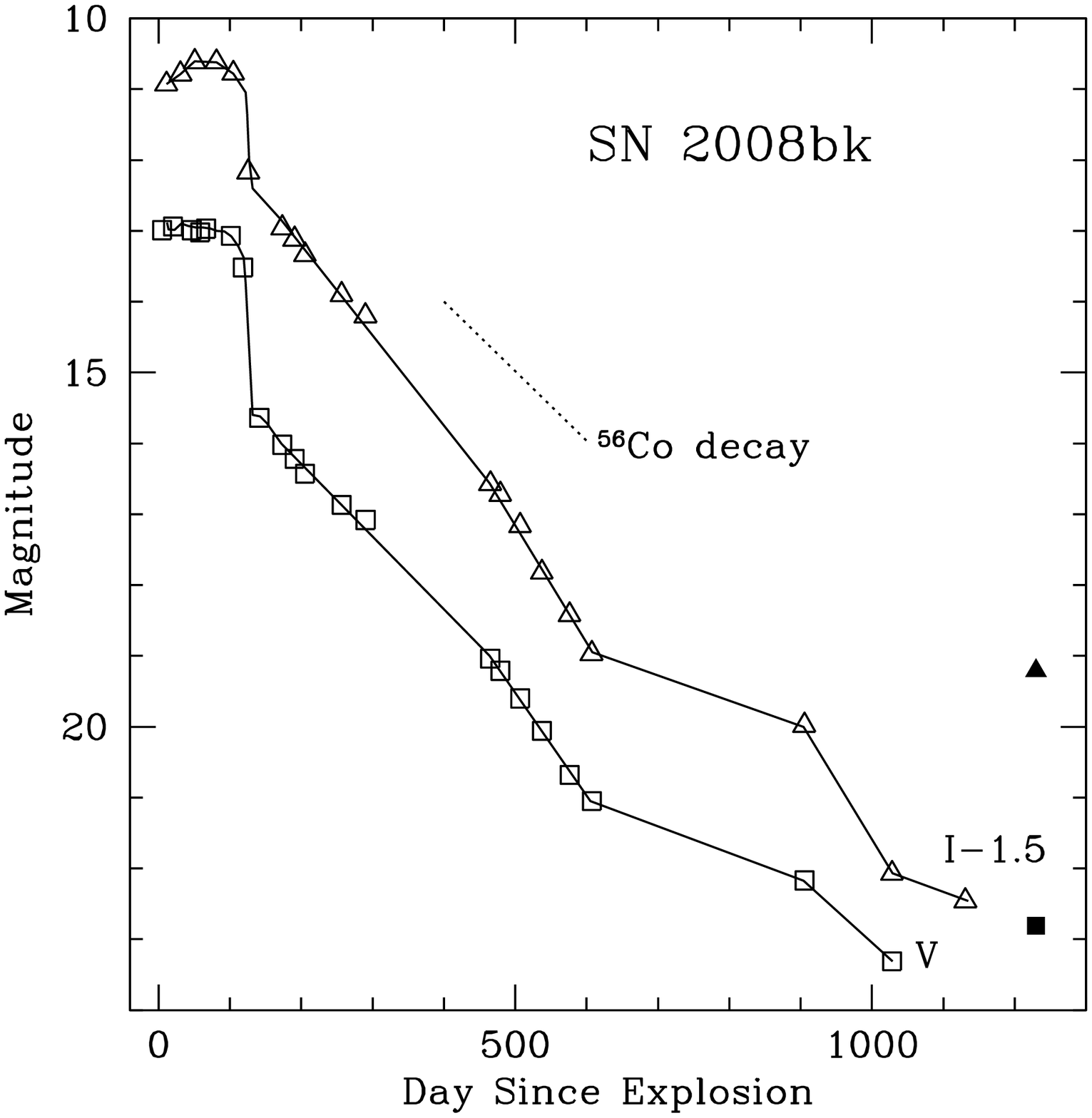}
\caption{Observed light curves in $V$ ({\it open squares}) and $I$ ({\it open triangles}) for SN 2008bk. 
Data shown are from \citet{van12a} and Table~\ref{tabphot}, except at $I$-band up to day $\sim$127,
which are preliminary values provided by G.~Pignata.
The $I$ light curve has been shifted by $-1.5$ mag for the sake
of clarity. The explosion date is assumed to be JD $\sim$2454548. 
For comparison at late times the expected decline rate associated with 
$^{56}$Co radioactive decay is shown ({\it dotted line}). The $V$ 
({\it solid square})
and $I$ ({\it solid triangle}, also shifted by $-1.5$ mag) measurements for the SN progenitor from 
\citet{van12a} are also shown for comparison with the light curves at the latest epochs.\label{figlc}}
\end{figure}

In addition, I include the earlier-time $V$ and $I$ photometry for SN 2008bk 
from \citet{van12a} and also preliminary $I$-band photometry up to  day $\sim$127
graciously provided by G.~Pignata.
The complete light curves in these two bands are shown in Figure~\ref{figlc}.

Shown for comparison in Figure~\ref{figlc} is the expected decline rate, 0.98 mag (100 d)$^{-1}$,
associated with thermalization by the ejecta of $\gamma$-rays from the radioactive decay of 
$^{56}$Co. The light curves on the tail 
essentially follow this decline rate, particularly at $V$ \citep[e.g.,][]{patat94}, up to day $\sim$450. 
However, after day $\sim$450 the
light curves decline more rapidly till day $\sim$600; this is likely due to the SN becoming more 
transparent to $\gamma$-rays, or possibly to the formation of dust, or both
\citep[e.g.,][]{hendry05}. After this time, the decline is more gradual up to
day $\sim$900, when the curves decline at a higher rate to reach what appears to be,
at least at $I$, a gradual decline once again. From a comparison with the late-time spectra of the
(peculiar) SN II-P 1987A in \citet{phillips90} and \citet{pun95}, one finds that the evolution of the 
luminosity in $V$ is likely driven primarily by changes in the relative strengths of nebular H$\alpha$, 
[O~{\sc i}]$\lambda\lambda$6300, 6364 and Na~{\sc i}~$\lambda\lambda$5890, 5896, whereas the 
$I$-band 
luminosity evolution is affected almost exclusively by 
[Ca~{\sc ii}]$\lambda\lambda$7291, 7323 and [Fe~{\sc ii}]$\lambda\lambda$7155, 7172 emission
(the [Ca~{\sc ii}] IR triplet and [C~{\sc i}]$\lambda$8727 line are present in this band, but are much 
weaker), which both fade in strength relative to H$\alpha$ after day $\sim$600. 

\section{Analysis of the Echo}\label{analysis}

I attempted to measure the brightness of the echo in both bands from the drizzled image mosaics.
Unfortunately, three, or possibly four, stars, or star-like objects, 
in the SN field are seen superposed along the arc.
I used PSF fitting in IRAF/DAOPHOT 
to subtract away the stars in the field, including most of those along the arc itself. 
Two of the stars would not subtract easily, so I let them remain in the
images, and then used the IRAF task ``listpix'' to obtain the individual pixel values in the remaining
portions of the arc, avoiding the residuals from the stars that did subtract away
and the profiles of the stars that did not.

I estimated the average count rate per pixel in the
echo to be $0.202 \pm 0.046$ e$^{-}$ s$^{-1}$ pixel$^{-1}$ in F606W  
and
$0.105 \pm 0.031$ e$^{-}$ s$^{-1}$ pixel$^{-1}$ in F814W over 72 pixels in each band. 
After
subtracting the average sky pixel count rate, 0.085 and 0.049 e$^{-}$ s$^{-1}$ pixel$^{-1}$ in F606W
and F814W, respectively, 
and assuming the VEGAMAG zero points from \citet{sirianni05} 
and a WFC plate scale of $0{\farcs}05$ pixel$^{-1}$,
these correspond to average surface brightnesses of 
$\textless{\mu}_{\rm F606W}\textgreater = 22.2 \pm 0.3$ 
and 
$\textless{\mu}_{\rm F814W}\textgreater = 22.1 \pm 0.5$ mag arcsec$^{-2}$.
Integrating in each band over the echo, which is an incomplete ring, 
I find
$m_{\rm F606W} = 23.7 \pm 0.4$ and $m_{\rm F814W} = 23.6 \pm 0.5$ mag,
with negligible difference resulting from the transformation (again, following \citeauthor{sirianni05}) 
to $V = 23.7 \pm 0.4$ 
and $I= 23.6 \pm 0.5$  mag, given the echo's color, i.e., $V-I=0.1 \pm 0.6$ mag.
Assuming Vega fluxes as the zero points, the echo has fluxes 
$F_{\rm echo}(V)=1.2 \pm 0.5 \times 10^{-18}$ and 
$F_{\rm echo}(I)=4.3 \pm 2.5 \times 10^{-19}$ erg cm$^{-2}$ s$^{-1}$ \AA$^{-1}$. 

The light echo ellipsoid can approximated as a paraboloid 
\citep[e.g.,][]{che86,schaefer87}.  
The perpendicular linear distance of
the line-of-sight to the SN from the line-of-sight to the echo 
is $b=D \theta$, where $D$ is the SN's distance from
Earth and $\theta$ is the angular distance between the two lines-of-sight.  
The echo width, as seen in the PSF-subtracted ACS F606W image, is unresolved (or barely
resolved), with a profile FWHM 
approximately that of point sources detected in this image before PSF subtraction. 
I measure the
echo's diameter to be 12.0 pixels from peak to peak of the profile across the echo. 
The echo radius, therefore, 
is 6.0 pixels, with an uncertainty of about $\pm 0.5$ pixel. This 
corresponds to
$\theta=0{\farcs}30 \pm 0{\farcs}03$.  Assuming the Cepheid distance to NGC 7793, 
$d=3.43 \pm 0.13$ Mpc \citep{piet10}, then $b=5.0 \pm 0.6$ pc.   

The distance from the SN to any scattering element along the echo,
$r=l+ct$, can be derived from $r^2 = b^2 + l^2$, where $l$ is the distance
from the SN to the echo along the line-of-sight \citep{couderc39}.  For $ct=0.86$ pc (2.81 ly), I 
find that $l= 14.0$ pc and $r=14.9$ pc, with an uncertainty  of $\pm 3.8$ pc, resulting from  
the uncertainties in the estimates of $b$ and $d$.

The echo is, therefore, most likely a result of scattering from interstellar, not circumstellar, dust, given
this distance. 
For a duration of the
RSG phase of $\sim 10^4$ yr and a wind speed $\sim 10$ km s$^{-1}$, the
circumstellar matter would only be $\sim 0.1$ pc ($0{\farcs}006$) in radius.
Additionally, much of any circumstellar  
dust was likely destroyed by the UV SN pulse \citep[e.g.,][]{sugerman03,van12b}.

It can  be assumed that the echo arises from single scattering in
a thin sheet of dust between us and the SN, and that the sheet thickness is
much smaller than the distance between the SN and the sheet (which is effectively true; see below).
The scattered flux $F$ at time $t$ from the echo at a given wavelength or bandpass is then 

\begin{equation}
F_{\rm echo} (t) = \int_0^t F_{\rm SN} (t-t^{\prime}) f(t^{\prime}) dt^{\prime},
\end{equation}

\noindent where $F_{\rm SN} (t-t^{\prime})$ is the fluence of the SN at time
$t-t^{\prime}$, and $f(t^{\prime})$, in units of inverse time, is the fraction of
light scattered by the echo toward the observer, which depends on the echo
geometry and the nature of the dust \citep{che86,cappellaro01,patat05}.  
The total SN light is effectively treated
as a short pulse over which the SN flux is constant, i.e., 
$F_{\rm SN} {\Delta}t_{\rm SN} = \int_0^{\infty} F_{\rm SN}(t) dt$ \citep{cappellaro01,patat05}.
The SN fluence is the integral of the light curves with respect to time in each band.  

Performing this integration (neglecting the overall small uncertainties in the observed SN  
photometry) and, again, assuming Vega as the flux zero point, I find 
$2.46 \times 10^{-7}$ and 
$1.90 \times 10^{-7}$ erg cm$^{-2}$ \AA$^{-1}$ in $V$ and $I$, respectively.
The duration of the SN pulse in each bandpass can be obtained by assuming
$F_{\rm SN}$ to be the SN maximum flux,
i.e., $V \approx 12.88$ and $I \approx 11.88$ mag.  
I then find ${\Delta}t_{\rm SN}$ to be
$\sim 108$ d in $V$ and $\sim 127$ d in $I$.
These durations are essentially equivalent to the plateau timescales in both bands, i.e., 
the light from the plateau, not surprisingly when the SN is
brightest, contributes the most to the overall SN pulse.
The effective pulse width \citep{sugerman02,sugerman03}, $w$, is then $\approx 0.9$ pc from $V$
and $\approx 1.0$ pc from $I$.
Since the echo is essentially unresolved, the quantity $\Delta b$ is less than the FWHM of a stellar
profile (or, $\sim$2 pixels), i.e., $\Delta b \lesssim 1.7$ pc.
The actual dust thickness from these observed quantities \citep[e.g.,][]{sugerman03} is 
then $\Delta l \lesssim 9.7$ pc.

In general, the term $f(t)$ is assumed to have the form \citep[e.g.,][]{cappellaro01,patat05} 

\begin{equation}
f(t)= {{c N_H} \over r} \int Q_{\rm sca}(a) \sigma_g(a) \Phi({\alpha}, a) \phi(a) da,
\end{equation}

\noindent 
where $N_H$ is the H number density, $Q_{\rm sca}(a)$ is the scattering
coefficient for a given grain radius $a$, $\sigma_g(a)= \pi a^2$ is the dust
grain cross section for scattering, and 

\begin{equation}
\Phi({\alpha}, a) = {{1 - g(a)^2} \over {4 \pi [1 + g(a)^2 - 2g(a) \cos(\alpha)]^{3/2}}},
\end{equation}

\noindent 
is the phase function \citep{hen41}, which applies for the bandpasses being 
considered here \citep{dra03}.
The term $g(a)$ measures the degree of forward scattering
for a dust grain of radius $a$. The scattering angle, $\alpha$, is 
defined by
$\cos(\alpha) = [(b/ct)^2 - 1] / [(b/ct)^2 + 1]$ \citep[e.g.,][]{schaefer87}.
 For this echo, 
the scattering angle is then $\alpha \approx 19{\fdg}6$. 
The term $\phi(a)$ is the grain size distribution for grain radius $a$.  Following
\citet{sugerman03}, 
I adopt the dust grain distributions for (spherical)
silicate and carbonaceous grains from \citet{weingartner01}. 
I assume ``standard'' diffuse interstellar dust and adopt the composition
of 53\% silicate and 47\% graphite grains from \citet{mrn77}, with the 
$Q_{\rm sca}(a)$ and $g(a)$ for ``smoothed UV astronomical silicate'' grains 
\citep{dra84,laor93,weingartner01} and for carbonaceous graphite \citep{dra84,laor93}.

\bibpunct[; ]{(}{)}{;}{a}{}{;}

I find, therefore, that I can fully account for the observed flux from the echo,
assuming that the SN light has been scattered by diffuse interstellar dust. The 
model results are $F_{\rm model}(V)=1.2 \pm 0.2 \times 10^{-18}$ and 
$F_{\rm model}(I)=4.3 \pm 1.0 \times 10^{-19}$ erg cm$^{-2}$ s$^{-1}$ \AA$^{-1}$ 
at $V$ and $I$, respectively, if I assume that $N_H=1.0 \times 10^{20}$ cm$^{-2}$ in the dust 
sheet producing the echo. 

The uncertainty in $F_{\rm model}$, above, formally arises from the uncertainty in $r$ (and also
therefore in $b$).
However, another source of uncertainty in the light echo profile is from the scattering dust width, 
which can range from fractions of a ly \citep[e.g., Cassiopeia A;][]{rest11b} to a
few ly \citep[e.g., SN 1987A;][]{sinnott13}, and can broaden the profile.
I have found, above, that the dust width is $\lesssim$32 ly; the calculation of $\Delta l$ is only
valid under the assumption that 
$\Delta l \gg w$, which appears to be essentially true based on the measured quantities.
Additional uncertainty arises from the inclination of the dust sheet with respect to the line of sight
\citep{rest11a,rest12a}, which can stretch the echo profile. 
The host galaxy itself has an inclination and position angle of $53{\fdg}7$ and
$99{\fdg}3$, respectively \citep{carignan90,bibby10}. 
The echo should be observed at a later epoch to measure the motion on the sky of the echo center 
and to infer the dust sheet inclination \citep{tylenda04,rest11a}.
(The echo did not appear to change perceptibly in shape or radius between the F814W images in 
2011 January and April.)
However, in the meantime I can provide an estimate of the inclination from the apparent offset in
the plane of the sky of the echo center from the SN position. Following \citet{tylenda04} for a 
plane-parallel dust slab, for an offset of 1.1 pc
and with the assumed value of $ct$, I find that 
an inclination angle of $\approx$52$\arcdeg$ is consistent with this measured offset.
Finally, the asymmetry in the SN explosion also introduces uncertainty in the echo properties 
\citep{rest11b,sinnott13}. For SN 2008bk it is unclear what was the degree of asymmetry in the 
explosion --- \citet{leonard12} found that the spectropolarimetric observations of this SN 
were not well accounted for by existing time-dependent radiative-transfer models for SNe II-P.
Nevertheless, the observed fluxes $F_{\rm echo}$ and model fluxes $F_{\rm model}$ that I have
calculated agree remarkably well, to within the formal uncertainties.

The relation between $N_H$ and visual extinction $A_V$, e.g., from \citet{guver09}, 
implies that the $A_V$ of the dust responsible for the echo is $\simeq 0.05$ mag.
This is interesting, in that \citet{van12a} assumed that the extinction within the
host galaxy immediately around SN 2008bk must be essentially zero, 
since a total extinction to the SN equivalent to that
from the Galactic foreground alone, i.e., $A_V=0.065$ mag from \citet{sch98}, 
was sufficient to account for the observed data.
A more recent estimate of the Galactic foreground extinction toward the SN is 
$A_V=0.054$ mag by \citet{schlafly11}\footnote{Using the foreground extinction calculator at
the NASA/IPAC Extragalactic Database, NED, at http://ned.ipac.caltech.edu/.}, allowing for
additional extinction of $\sim 0.01$ mag to exist internal to the host galaxy along the line-of-sight
to the SN.
The echo, though, must result from a denser part of the dust sheet than what we peer through 
toward the SN.
This is consistent with the filamentary nature of dust in the interstellar medium 
and its effect on observed light echoes \citep[e.g.,][]{rest11a}. 
One can see the non-uniform surface brightness of the
echo in Figure~\ref{figecho}, which likely results from filaments within the
dust sheet. For an echo at 3.4 Mpc, even with {\sl HST}, we do not have the luxury of 
the precise spatial detail seen in the dust structures
responsible for the echoes around the much closer Cas A \citep[e.g.,][]{rest08b} 
or SN 1987A \citep[e.g.,][]{crotts88}, 
and therefore cannot 
fully model the actual dust distribution in the environment of SN 2008bk.

\section{Discussion and Conclusions}\label{discussion}

The SN 2008bk progenitor was identified in ground-based images, albeit of high quality
(seeing $\sim 0{\farcs}65$ FWHM for the Gemini images, \citeauthor{van12a}~\citeyear{van12a};
$0{\farcs}5$, $0{\farcs}8$, and $0{\farcs}4$ FWHM for the $JHK_s$ VLT images, 
\citeauthor{mat08b}~\citeyear{mat08b}).
With the SN now having significantly faded, several much fainter stars around it are revealed 
in the {\sl HST\/} images (see Figure~\ref{figecho}). This, then, raises the question as to 
what level these stars contributed to the overall brightness of the progenitor star, as seen 
from the ground. The processing of the {\sl HST\/}
data within Dolphot also resulted in photometry for these stars, which is given in 
Table~\ref{tabstars}.
In the first place, the stars are all at a level of a factor $\lesssim 0.15$ (15\% or less) 
of the progenitor flux at $V$, 
$\lesssim 0.39$ at $I$ (the brightest is Star A, which is $\sim 0{\farcs}45$ from the SN),
and $\lesssim 0.10$ at $J$ and $H$. All of these stars are, of course, in the wings of
the progenitor star's profile, where the actual fractional contribution of their fluxes to that of
the progenitor would be further diminished.

I have analyzed each of the pre-SN Gemini-S GMOS and VLT images considered by \citet{van12a}, 
carefully examining the residual images produced by the DAOPHOT PSF fitting (and
subsequent subtraction) used to obtain the stellar photometry.
As best as I could discern, by comparing the {\sl HST\/} images to the ground-based residual 
images, the stars from Table~\ref{tabstars}, when they actually could be distinguished 
(which was generally only for the brighter of these stars in each band), appear to persist after PSF 
subtraction.
In other words, DAOPHOT had adequately fit the profile of the progenitor star in each of the
ground-based images, leaving the surrounding stars as part of the residual from the fit.
In particular, stars A through D are clearly seen in the Gemini $i'$-band residual image. All of the stars
are comparatively fainter at $V$, so it is not certain that they would have been well detected in
the Gemini $g'$ image before PSF fitting. Similarly, the $J$ and $K_s$ VLT images actually 
may not have been deep enough to detect these faint stars, compared to the much brighter progenitor.
I therefore conclude that these stars likely had a negligible effect on the progenitor's brightness in all 
of the bands, as seen from the ground. 
I note that, as the SN continues to fade, other stars may appear within the SN's, and, therefore, 
the progenitor's, PSF. (I have already noted, above, that the SN PSF is asymmetric at $V$, possibly
due to a very close neighboring star.)
However, the overall satisfactory comparison that \citet{van12a} made
between the model spectral energy distribution and the observed photometry for the progenitor is 
consistent with an overall lack of significant contamination by immediate neighbors.

\bibpunct[; ]{(}{)}{;}{a}{}{;}
In summary, I have discovered a prominent light echo around the SN II-P 2008bk in NGC 7793 in 
archival {\sl HST\/} images obtained at late times.
The echo can be fully explained by the scattering of the observed light curves for the SN by a
distribution of standard interstellar dust at a distance of $\approx 15$ pc from the SN. The density
of the echo-producing dust is larger than that toward the SN itself, since the inferred extinction of
the echo dust is $A_V \approx 0.05$ mag, whereas the extinction internal to the host toward 
the SN is much lower, $A_V \approx 0.01$ mag. 
Additionally, from the very late-time light curves it is evident that the progenitor star has vanished.
The stars in the immediate environment of SN 2008bk do not appear to have contaminated the 
photometry of the progenitor, and therefore the estimates of the bolometric luminosity, 
effective temperature, and initial mass made by \citet{van12a} for the star continue to hold.

This echo should be further monitored with {\sl HST}, particularly in the blue, 
to better constrain the nature of the scattering dust and the echo geometry, and to reveal further 
new or evolving structures in the echo.

\acknowledgments

I appreciate the careful review by the referee and am grateful 
for the helpful comments provided that improved this paper.
I am also grateful to Giuliano Pignata for providing preliminary early-time photometry of the SN at 
$I$-band.
This work was based in part on observations made with the NASA/ESA
{\it Hubble Space Telescope}, obtained from the Data Archive at the
Space Telescope Science Institute, which is operated by the
Association of Universities for Research in Astronomy (AURA), Inc.,
under NASA contract NAS 05-26555.
It was also based 
in part on data obtained from the ESO Science Archive Facility under request numbers
8354, 31942, and 31948.

\begin{deluxetable}{lcccccccc}
\tablewidth{0pt}
\tablecolumns{9}
\tablecaption{Late-Time Photometry of SN 2008bk\tablenotemark{a}\label{tabphot}}
\tablehead{
\colhead{UT date} &\colhead{JD$-$2450000} & \colhead{F555W} &
\colhead{$V$} & \colhead{F814W} & \colhead{$I$} & 
\colhead{F125W} & \colhead{F160W} & \colhead{Source\tablenotemark{b}} \\
\colhead{} &\colhead{ } & \colhead{(mag)} &
\colhead{(mag)} & \colhead{(mag)} & \colhead{(mag)} & 
\colhead{(mag)} & \colhead{(mag)} & \colhead{}}
\startdata
2009 Jan 09.55 & 4840.05 & \nodata & 17.08(04) & \nodata & 15.70(02) & 
\nodata & \nodata & NTT/EFOSC2 \\
2009 Sep 05.71 & 5080.21 & \nodata & 20.10(04) & \nodata & \nodata & 
\nodata & \nodata & VLT/FORS2 \\
2009 Sep 13.67 & 5087.17 & \nodata & 20.06(04) & \nodata & 19.32(03) & 
\nodata & \nodata & NTT/EFOSC2 \\
2009 Oct 22.52 & 5126.02 & \nodata & 20.68(04) & \nodata & 19.92(04) & 
\nodata & \nodata & NTT/EFOSC2 \\
2009 Nov 22.55 & 5157.05 & \nodata & 21.05(04) & \nodata & 20.47(04) & 
\nodata & \nodata & NTT/EFOSC2 \\
2010 Sep 16.69 & 5455.19 & \nodata & 22.17(07) & \nodata & 21.48(12) & 
\nodata & \nodata & NTT/EFOSC2 \\
2011 Jan 17.73 & 5578.23  & 23.35(01) & 23.31(01) & 23.58(02) & 23.57(02) & 
\nodata & \nodata & {\sl HST}/ACS \\
2011 Apr 29.48 & 5680.98 & \nodata & \nodata & 23.96(05) & &
23.68(08) & 23.78(10) & {\sl HST}/WFC3 \\
\enddata
\tablenotetext{a}{Uncertainties $(1\sigma)$ are given in parentheses, in
units of hundredths of a mag.}
\tablenotetext{b}{NTT/EFOSC2 = New Technology Telescope + 
ESO Faint Object Spectrograph and Camera (EFOSC2);
VLT/FORS2 = Very Large Telescope + FOcal Reducer and low dispersion Spectrograph (FORS2);
{\sl HST}/ACS = {\sl HST\/} + Advanced Camera for Surveys Wide-Field Channel;
{\sl HST}/WFC3 = {\sl HST\/} + Wide-Field Camera 3 UVIS/IR.}
\end{deluxetable}

\begin{deluxetable}{ccccccccc}
\tablenum{2}
\tablewidth{6.9truein}
\tablecolumns{9}
\tablecaption{Stars Around SN 2008bk\tablenotemark{a}\label{tabstars}}
\tablehead{
\colhead{Star} &\colhead{$\Delta$\tablenotemark{b}} & \colhead{F606W} &
\colhead{$V$} & \colhead{ACS/F814W} & \colhead{$I$} &
\colhead{WFC3/F814W} & 
\colhead{F125W} & \colhead{F160W} \\
\colhead{} & \colhead{($\arcsec$)} & \colhead{(mag)} &
\colhead{(mag)} & \colhead{(mag)} & \colhead{(mag)} &
\colhead{(mag)} & 
\colhead{(mag)} & \colhead{(mag)}
}
\startdata
A & 0.45 & 24.65(03) & 25.12 & 23.05(01) & 23.04 & 23.09(03) & 21.80(02) & 21.11(01) \\
B & 0.47 & 25.47(05) & 25.89 & 24.13(03) & 24.02 & 24.35(06) & 22.89(04) & 22.36(03) \\
C & 0.28 & 25.18(04) & 25.57 & 23.93(03) & 23.94 & 24.17(06) & 22.78(04) & 22.15(03) \\
D & 0.55 & 25.25(04) & 25.63 & 24.05(03) & 24.13 & 24.32(06) & 23.13(05) & 22.41(03) \\
E & 0.54 & 25.36(04) & 25.36 & 25.35(07) & 24.98 & 25.52(15) & \nodata & \nodata \\
F & 0.28 & 25.01(03) & 24.98 & 25.21(06) & 25.09 & 25.58(15) & 22.89(04) & 22.36(03) \\
G & 0.38 & 26.04(07) & 26.21 & 25.48(08) & 25.35 & 25.41(13)    & \nodata & \nodata \\
\enddata
\tablenotetext{a}{Uncertainties $(1\sigma)$ are given in parentheses, in
units of hundredths of a mag.}
\tablenotetext{b}{This is the total offset of the centroid of the star from that of the SN.}
\end{deluxetable}

\clearpage

\begin{figure}
\includegraphics[angle=0,scale=0.70]{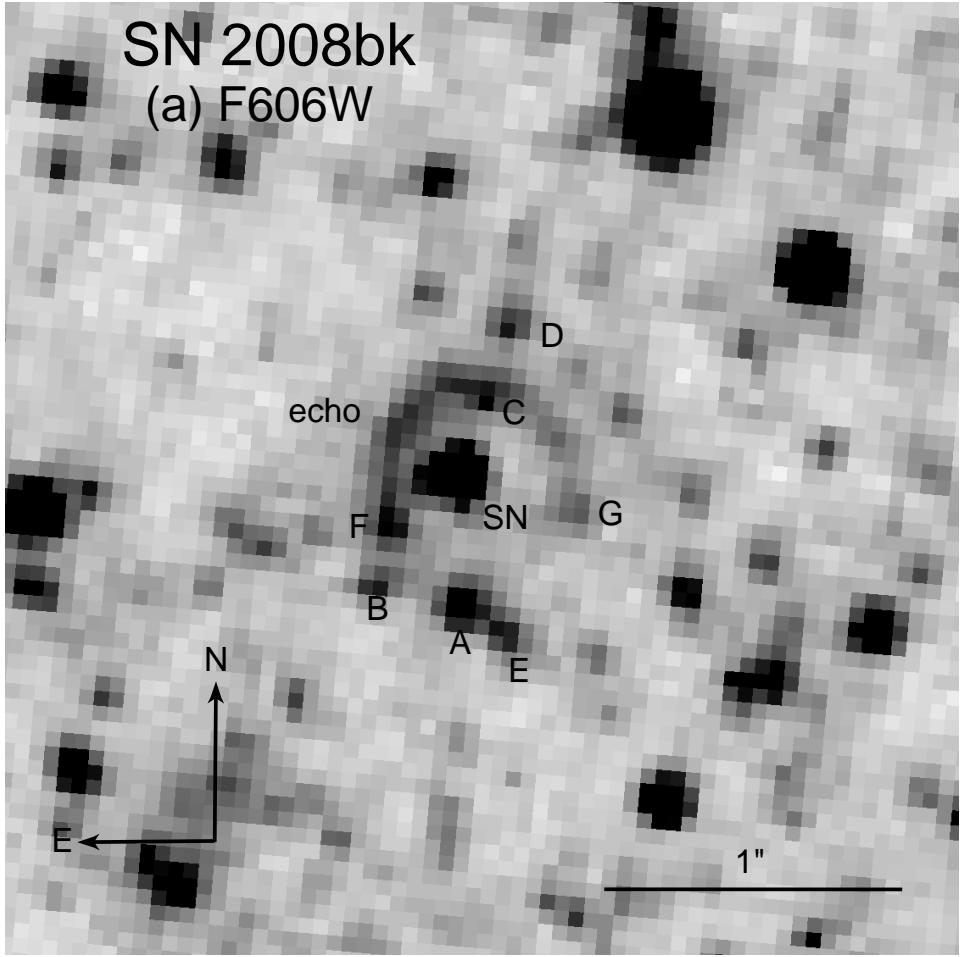}
\caption{A portion of the archival {\sl HST\/} images obtained with the ACS/WFC, containing the site 
of the SN II-P 2008bk in NGC 7793, at late times (SN age $\sim 1024.7$ d, or 2.81 yr) in
the ({\it a}) F606W ($\sim V$) and ({\it b}) F814W ($\sim I$) passbands.
A prominent light echo is visible around the SN. 
Stars within $\sim 0{\farcs}6$ of the SN are also indicated in the figure; see Table \ref{tabstars}.
North is up, and east is to the left.\label{figecho}}
\end{figure}

\clearpage

\begin{figure}
\figurenum{1}
\includegraphics[angle=0,scale=0.70]{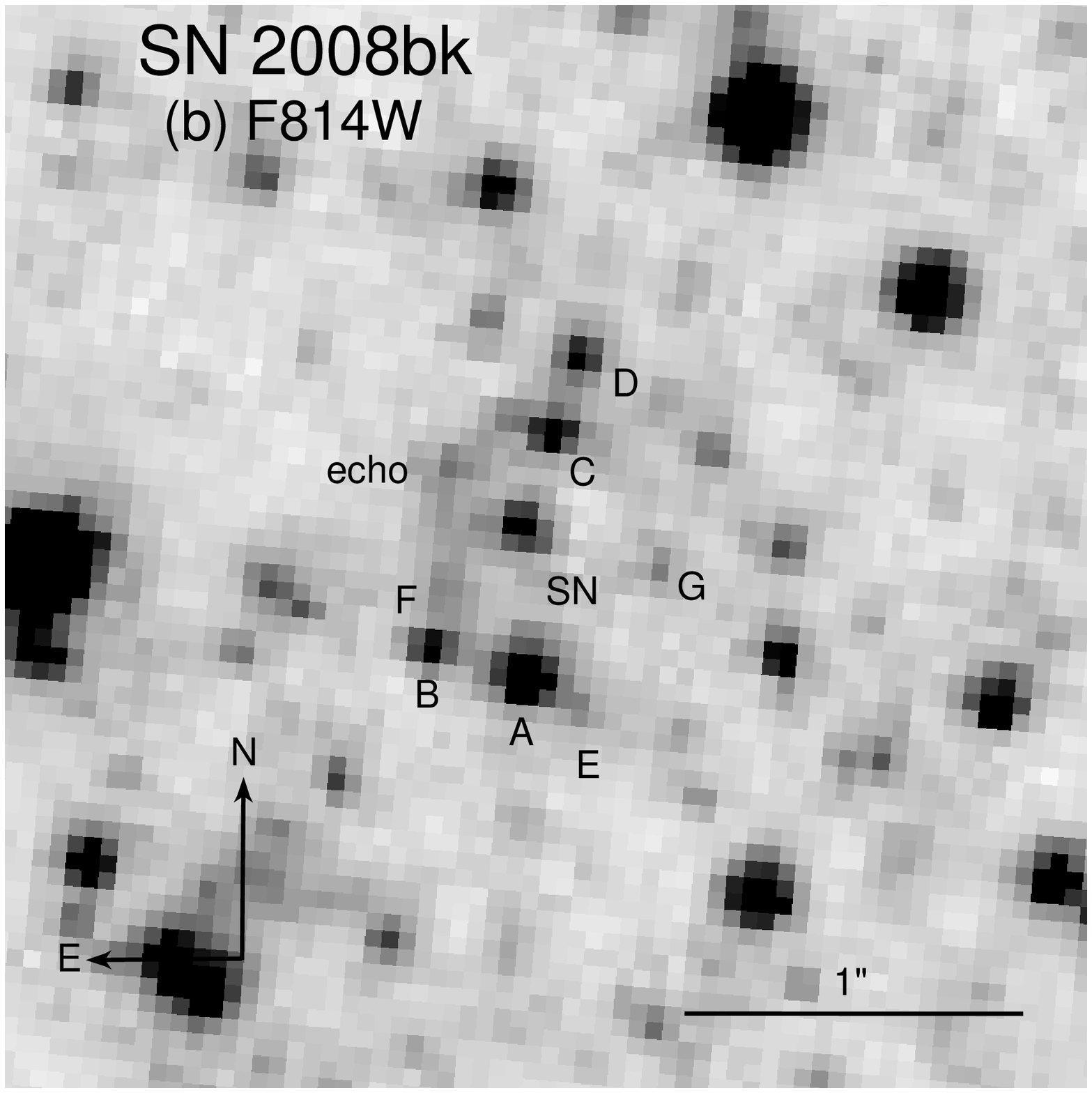}
\caption{(Continued.)}
\end{figure}

\clearpage

\end{document}